\documentclass[12pt]{article}
\usepackage{amsfonts}
\usepackage{amssymb}

\def\beq{\begin{equation}}
\def\eeq{\end{equation}}
\def\bea{\begin{eqnarray}}
\def\eea{\end{eqnarray}}
\def\bq{\begin{quote}}
\def\eq{\end{quote}}

\def\ba{\begin{array}}
\def\ea{\end{array}}

\begin{document}
\date{}
\title{Interpreting the lattice monopoles\\
in the continuum terms
\vskip-40mm
\rightline{\small ITEP-LAT/2002-25}
\rightline{\small MPI-PhT/2002-68}
\vskip 40mm
}

\author{F.V. Gubarev \\
{\small Institute of Theoretical and  Experimental Physics,}\\
{\small B.~Cheremushkinskaja 25, Moscow, 117259, Russia}\\
\\
V.I.~Zakharov \\
{\small Max-Planck Institut f\"ur Physik,}\\
{\small F\"ohringer Ring 6, 80805, M\"unchen, Germany}
}

\maketitle
\begin{abstract}\noindent
We review briefly current interpretation of the lattice monopoles, defined within the Maximal
Abelian Projection, in terms of the continuum theory. We emphasize, in particular, that
the lattice data, at the presently available lattices, indicate that the monopoles are
associated with singular fields. This note is prompted by a recent analysis hep-lat/0211005 
which is based on an implicit assumption that the fields are regular.
\end{abstract}

\section{Introduction}

Writing these notes was originally motivated by the recent
paper ``On Nonexistence of Magnetic Charge in Pure Yang-Mills
theories'' \cite{kovner}. The paper appears critical about a few
earlier papers written with our participation and our intention was
to appreciate the critique and summarize our (respectively modified)
views. However, a closer reading of the paper in Ref.~\cite{kovner}
reveals that the points which were central
for us are not even mentioned in \cite{kovner}.
In particular, the paper in Ref.~\cite{bornyakov01}
reports the following result:
\beq
\label{access}
S^{lat}_{mon}~-~S^{lat}_{av}~=~f(a) \qquad \mathrm{[lattice ~ units]}\,,
\eeq
where $S^{lat}_{av}$ is the average action per plaquette over the whole
lattice, $S^{lat}_{mon}$ is the action on the plaquettes closest to the 
position of the monopoles (occupying center of a cube) and $f(a)$
is a positive function which depends slightly on the lattice spacing $a$
for the lattices tested. Note that Eq.~(\ref{access}) is written in lattice
units. Moreover the monopoles themselves are defined within 
the Abelian-projected theory while $S^{lat}_{av,mon}$ is
{\it the full non-Abelian action} of the original lattice SU(2) gluodynamics. 

The fact that the access of the action associated with the monopoles
is finite in the lattice units implies that at the presently available 
lattices:   
\beq
|(G_{\mu\nu}^a)_{mon}|~\sim ~{1\over a^2}\,,
\eeq
where $(G_{\mu\nu}^a)_{mon}$ is the non-Abelian field-strength tensor 
associated with the monopole. In other words, in the limit
$a\to 0$ the monopole fields appear  singular
(if we extrapolate the observed behavior to $a\to 0$ limit\footnote{
Through the whole paper we actually mean only the lattice
results valid  on the available lattices and consider $1/a$ as the 
ultraviolet (UV) cut off. We do not speculate here
what actually happens or should happen at lattice spacings 
much smaller than those tested. Thus, if we simply say, 
for example, that a quantity is UV divergent this 
actually assumes ``if the presently observed dependence on $a$
holds for $a\to 0$''. Since we are going to discuss only interpretation
of the existing data this reservation is not important for our purposes.
}) and this is the main implication of \cite{bornyakov01}.
On the other hand, there is no single line in  \cite{kovner}
which mentions singular fields. The critique \cite{kovner} is written
in a {\it bona fide} conviction that the authors know the content of
the papers criticized without examining them.

Although we cannot compare our views with those of the authors 
of \cite{kovner} directly for this reason, we could view the paper \cite{kovner}
in a more general context as a reminder 
that it is better to have interpretation of the lattice data
in terms of the continuum theory.
Finding such an interpretation is, indeed,  a long overdue and a real 
challenge, see, e.g.,~\cite{alive}. 
Moreover, saying that the monopole fields
are actually singular is no panacea at all and in fact makes  the challenge even
more acute. Thus, we decided to summarize the current interpretation
of the lattice data on the monopoles, as we understand it. There are many
open ends yet and there is little doubt
that our current understanding misses some important points.
However, we find the interplay between the lattice and continuum
exciting and productive. 

One more remark is in order now. While the authors of \cite{kovner}
discuss (non)existence of the monopoles we prefer to discuss 
interpretation of the data. It might well be just a matter of language, but the
very observation (\ref{access}) makes the problem of existence of
the monopoles redundant. They exist so to say by definition since have
a non-vanishing gauge-invariant action\footnote{
The authors of \cite{kovner} might have argued that their non-existence theorem
implies the data (\ref{access}) to be wrong. It is worth mentioning therefore
that non-vanishing of the r.h.s. of (\ref{access}) in the lattice units 
can be checked against quite a few publications, see, in particular,
\cite{many}. Ref.~\cite{bornyakov01} addresses mainly more subtle
points to be mentioned below.
}. However, theoretical interpretation  of the results is of course an open
issue. In particular, they might not be the ``true'' monopoles, whatever it means. 
They cannot be, however, gauge artifacts.

The organization of the paper is as follows. In Sect.~2 we review briefly
the 't Hooft's definition of the monopoles \cite{thooft}
and its implementation on the lattice. As far as the case of regular fields
is concerned, our conclusions are close, albeit not identical,
to those of Ref.~\cite{kovner}. In Sect.~3 we turn to the realistic case of singular
fields and review data on fine tuning of the monopoles. In Sect.~4 we confront some further
theoretical problems brought in by the fine tuning.

\section{Defining monopoles in pure gauge theories}

\subsection{'t Hooft`s procedure}

Generically, all the definitions of the monopoles on the lattice 
can be traced back to the seminal paper by 't Hooft \cite{thooft}.
For simplicity we will consider only the $SU(2)$ case. Then one
introduces a color vector $\chi^a~ (a=1,2,3)$ and partially fixes the
gauge by rotating the vector to the third direction in the color space.
This leaves $U(1)$ gauge freedom untouched since the gauge rotations
around the third axis are still allowed. Then one can demonstrate that
the world-lines where all the three components of $\chi^a$ vanish,
\beq
\chi^a~=~0\,, \quad a=1,2,3\,,
\eeq
are in fact monopole trajectories where the monopole
charge is defined with respect to the $U(1)$ group which is the remaining gauge
freedom, see above. Moreover, the magnetic flux can be evaluated
in terms of the 't Hooft tensor $F_{\mu\nu}$:
\beq\label{fmunu}
F_{\mu\nu}~=~G_{\mu\nu}^a \, n^a ~+~ \epsilon^{abc}\,
n^a\,D_{\mu}n^b\,D_{\nu}n^c\,,
\eeq
where $n^a~\equiv~\chi^a/|\chi^a|$. Note that the field $\chi^a$
in many respect plays the role of the standard Higgs field
of the Georgi-Glashow model.

This procedure fixes uniquely a definition of a monopole and its
position is gauge invariant since vanishing of
a color vector is a gauge invariant condition. However, 
the interaction of such monopoles between themselves and with,
say, Wilson loop is an open question since $F_{\mu\nu}$ does not enter
directly the action. Moreover, unlike the Georgi-Glashow model,
the vacuum expectation of $\chi^a$ is to vanish,
\beq \label{zeros}
\langle \chi^a \rangle~=~0\,,
\eeq
because of the color conservation. The best that one can do to link
the dynamics of the monopoles defined in terms of (\ref{fmunu})
to the standard dynamics of the Georgi-Glashow model is to assume 
a domain structure such that $|\chi^a|\approx const$ within a domain.
As far as we know, neither the domain picture nor
the mechanism of the Abelian dominance in terms
of the monopoles defined this way has ever been analyzed in any detail. 

\subsection{C-parity}

In case of $SU(2)$ gluodynamics the simplest composite 
(pseudoscalar) field $\chi^a$
is of the form:
\beq\label{three}
\chi^a~=~ \epsilon^{ade}\,\epsilon^{bcd}\,\epsilon_{\mu\nu\lambda\rho}\,\,
G^b_{\mu\alpha}\,G^c_{\nu\alpha}\,G^e_{\rho\lambda}\, ,
\eeq
while the simplest scalar field can be constructed only on four 
gluonic field-strength tensors.

Moreover, the  field (\ref{three}) has a negative C-parity,
\beq\label{sigma}
C(\chi^3)~=~-1\,,
\eeq
while the C-parity of the standard Higgs field of
the Georgi-Glashow model is positive.
This observation about the $C$-parity 
can be generalized on any composite field $\chi^a$
in the $SU(2)$ case. Indeed, $G$-parity of the triplet of the gluon fields is
positive while the G-parity of the standard Higgs is negative. 
This is the central point of the paper \cite{kovner} which concludes that
in the SU(2) case monopoles cannot be defined.

While the theorem that all the composite $\chi^a$ fields 
have positive G-parity is certainly true, we are less sure that
it precludes us from using the composite fields
for the monopole definition as a matter of principle.
Indeed, usually, the G-parity of the Higgs field is fixed
by the condition that the final theory should conserve C-parity
and by the observation that the v.e.v. $<\chi^3>\neq 0$.
Since in our case the v.e.v. vanishes anyhow (see (\ref{zeros}))
the G-parity of the effective Higgs field might be not fixed either.
There is no discussion of this point in~\cite{kovner}.
From the overview above it does not follow, we feel, that the
field $\chi^a$ should have negative G-parity.  
However, we would not discuss the issue in detail  either 
since the very possibility that a composite  Higgs plays 
a significant dynamical role seems to us to be too remote
from the physics of the lattice monopoles. 

\subsection{Maximal Abelian Projection}

By ``monopoles'', without further specifications, we mean in this
note the monopoles of the Maximal Abelian Projection (MAP),
for review see, e.g., \cite{reviews}. To define the monopoles, 
one fixes first the maximal Abelian gauge by minimizing
\beq
R~=~\sum\limits_{l}\,[(A^1_l)^2 + (A^2_l)^2]\,, 
\eeq
where the sum is taken over all links on the lattice and the 
upper index of the gauge potential indicates the color.
As usual for the procedures following the pattern proposed in \cite{thooft},
the gauge is fixed up to U(1) rotations around the third axis.

The MA projection is defined then by the putting 
$A^{1,2}_l\equiv 0$. Finally, the MAP monopoles are the monopoles 
associated with $A^3_l$.

Locally and in the continuum limit the procedure described 
corresponds to choosing the gauge
\beq
\big(\partial_{\mu} \mp igA_{\mu}^3\big)A^{\pm}_{\mu}~=~0\,.
\eeq
Note that we do not introduce any composite Higgs field now.
The fixation of the gauge can fail only because of the singularities
in $A_{\mu}^3$. As usual, the lines where the fixation 
of the gauge fails are associated with the monopoles which are
the topological defects in MA projected theory.

The magnetic current $j_{\nu}$
is defined in terms of violations of the Bianchi
identity:
\beq\label{bianchi}
\partial_{\mu}\tilde{F}_{\mu\nu}~\equiv~j_{\nu}~,
\eeq
where $\tilde{F}_{\mu\nu}\equiv 1/2 \, \epsilon_{\mu\nu\rho\sigma}F_{\rho\sigma}$.
By the field strength one understands here the $F_{\mu\nu}$ of the projected
theory. Moreover, since violations of the Bianchi identity assume
singular field, Eq.~(\ref{bianchi}) is to be understood as
regularized on the lattice \cite{degrand}.

There are no problems with the C-parity of the current defined through
(\ref{bianchi}) -- the issue of concern for the authors of \cite{kovner}.

Thus, it is easy to argue that the MAP monopoles are associated with
singular gauge potentials. This is an alternative to introducing 
composite Higgs fields. However, one might have thought that this 
singularity in the (projected) potential is not related to any
original, non-Abelian  action. The data (\ref{access}) show that 
in fact there is a singularity in the non-Abelian action as well.
To accommodate for such a possibility we should readjust
considerably our way of thinking.

\section{Fine tuning}

\subsection{Tuning the action and entropy factors}

There are quite a few reasons why one would be inclined 
to disregard singular fields. First of all, apparently,
such fields are associated with an infinite action.
Indeed, if the action density associated with the monopoles, see 
(\ref{access}), is in the lattice units the field-theoretical
mass of the monopole is also in the lattice units, that is:
\beq
M_{mon}(a)~\sim~\int d^3r ({\bf B^a})^2~\sim ~{1\over a}\,,
\eeq
where the lattice spacing plays the role of the UV cut off in the
continuum limit, ${\bf B^a}$ is the color magnetic field.
Thus, the action suppression 
is infinitely strong in the limit $a\to 0$:
\beq
e^{-S}~\sim~e^{-\mathrm{const} \cdot L/a}\,,
\eeq
where $L$ is the length of the monopole trajectory.
At presently available lattices one gets an estimate
for the monopole mass:
\beq
M_{mon}~\gtrsim~5~\mathrm{GeV}\,,
\eeq
which corresponds to the smallest available $a ~ \approx ~ 0.06 ~ \mathrm{fm}$.

Naively, excitations with such mass would be mere lattice artifacts.
However, the monopole trajectories have scaling properties. 
Without reviewing all the data, let us mention that in the confining
phase there is a single percolating cluster which fills in the whole
of the lattice. Moreover, there are self-intersections 
in the percolating cluster. Associating  intersections with interactions, 
one  can introduce a ``free-path length'' of the monopole, $L_{free}$ 
as the length of the segments of the trajectories between the intersections.
According to the measurements \cite{boyko} $L_{free}$ scales
and does not depend on $a$:
\beq\label{free}
L_{free}~\approx~1.6~\mathrm{fm}\,.
\eeq
In other words:
\beq\label{paradox}
L_{free} \cdot M_{mon} ~\gtrsim ~ 40\,,
\eeq
where the numerical factor on the r.h.s. is related to the smallest $a$ available.

We are in haste to add that Eq.~(\ref{paradox}) is in no contradiction with
Quantum Mechanics. The point is that $M_{mon}(a)$ entering the action, $S_{mon}=M_{mon}\cdot L$, is
{\bf not} the propagating mass in the lattice formulation of the 
field theory. For a free particle the propagating mass $m_{prop}$ is related to it as:
\beq\label{ft}
m^2_{prop}\cdot a ~ \approx ~ 8\big[\, M_{mon}(a)\,-\,{\ln 7\over a}\,\big]\,.
\eeq
The factor proportional to the $\ln 7$ is due to the entropy,
i.e. to the number of trajectories of the same length $L$.
Derivations of (\ref{ft}) can be found in textbooks, see, e.g.~\cite{ambjorn}.
(It might worth noting that $\ln 7$ is specific for the monopoles on the cubic
lattice). 

According to the direct measurements \cite{bornyakov01}
the monopole action is indeed close to $S=\ln 7\cdot L/a$.
However, for a number of reasons it is difficult to measure directly
how fine the tuning is. Indirect evidences, like (\ref{paradox}),
are more powerful.    

Thus, the lattice data  strongly indicate \cite{vz} 
that the monopoles are {\bf fine-tuned} objects with a huge
cancellation between the suppressing factor due to the action and 
enhancing factor due to the entropy. Without taking this fine tuning into 
consideration, any discussion of the physics of the
lattice monopoles is abstract.

\subsection{Gauge-invariant ``definition'' of the monopole}

Since the fine tuning is formulated in terms of the non-Abelian action,
one could use it to {\it define} the monopoles\footnote{
A similar remark was made independently by T. Suzuki.
}. Namely, the monopole currents could be defined as closed trajectories
along which the average action is fine tuned to the entropy factor.
Note that the entropy factor would change with a change of the lattice.

Of course such a definition can be discussed only as a matter of principle.
For an individual plaquette associated with a monopole 
the action can be far from its average value. Indeed, we deal with a 
quantum process of measurement.

\section{Implications of the fine tuning}

Realization that the lattice monopoles are fine tuned 
brings actually further difficult questions. First of all, 
the fine tuning above was discussed in terms of light particles,
but this cannot be true -- everybody feels this way -- that
we have new light particles. Since the monopole fine tuning
is quite a recent notion, the answers
to this and other questions are very provisional at best. 
Nevertheless we would discuss, in an exploratory way,
a few further points.

\subsection{Two facets of the fine tuning}

Now, we turn to the question whether the fine tuning 
implies introduction of new light particles
(and is unacceptable for this reason). A simple argument
shows that it is not necessarily so \cite{vz}.

Indeed, imagine that (see Eq.~(\ref{ft}))
\beq
\label{imaging}
M_{mon}(a)~-{\ln 7\over a}~\sim ~ \Lambda_{QCD}\,.
\eeq
This kind of fine tuning is suggested by the scaling of $L_{free}$, see
Eqs.~(\ref{free},\ref{paradox}). On the other hand, the corresponding
``mass'' of the monopole is then of order
\beq\label{infty}
m^2_{prop}~\sim~{\Lambda_{QCD}\over a}\,,
\eeq
and the monopoles are removed from the spectrum in the limit 
$a\to 0$.

There is another relation which indicates that interpretation of the
monopoles in terms of a magnetically charged field reveals a
non-standard picture. The point is that the density of the
monopoles in the percolating cluster scales:
\beq
\rho_{perc}~\sim ~\Lambda^{-3}_{QCD}\,,
\eeq
see in particular \cite{bornyakov02}. Here  $\rho_{perc}$ is
defined in terms of the length of the percolating cluster:
\beq
L~=~\rho_{perc}V_4\,,
\eeq
where $V_4$ is the 4-volume of the lattice. If one considers the percolating  cluster
of the  monopoles as representing the ground state of the condensed monopoles, then 
\beq\label{zero}
\langle |\phi|^2\rangle~\sim~\rho_{perc}\cdot a\,,
\eeq
where $\phi$ is the magnetically charged field. 

According to (\ref{zero}) the condensate $\langle |\phi|^2\rangle$
vanishes in the limit $a\to 0$. However, the vacuum expectation
value of the product $m^2|\phi|^2$ scales, compare (\ref{infty}).
Generalizing this observation one can speculate that the vacuum
expectation value of the whole potential energy scales similarly:
\beq\label{energy}
\langle V(|\phi|^2)\rangle~\sim~\Lambda_{QCD}^4\,.
\eeq
Eq.~(\ref{energy}), if confirmed by further analysis, would
make manifest the relevance of the field $\phi$ to the non-perturbative
QCD.

We pause here to emphasize a two-facet nature of the monopole fine tuning.
We do introduce singular fields (\ref{ft}) but this is necessary to
cancel the singular entropy factor. The singularity is
specific for the Euclidean (and lattice) formulation.
If, for example, the geometry
of the lattice is changed, the singular fields are predicted to change 
as well, in such a way as to maintain the cancellation with the entropy
factor. Eq.~(\ref{energy}) reveals this facet of the 
fine tuning even more convincingly.  Namely, one could postulate 
from the very beginning that the vacuum expectation value of the magnetically
charged field cannot not diverge in the ultraviolet. This is now a manifestation, 
in the field theoretical language, of the fact that $\phi$ is
an effective field describing non-perturbative effects
(although in the language of fine tuning it is associated with
singular original fields).

\subsection{Monopole current}

If we introduce singular fields in the continuum limit,
the monopole current $j^a_{\nu}$ is naturally defined in terms
of the fields themselves:
\beq\label{current}
(D_{\mu}\tilde{G}_{\mu\nu})^a~=~j^a_{\nu}\,.
\eeq
In other words the monopole current is associated with
the violation of the Bianchi identity which may arise
because of the singularities in the fields. It goes without
saying that (\ref{current}) is rather postulated than derived.
Unfortunately, there exists no helpful lattice implementation of Eq.~(\ref{current}).
Therefore, one cannot actually check that the definition (\ref{current}) corresponds to the
monopoles observed on the lattice, although there is a little doubt
that Eqs.~(\ref{bianchi},\ref{current}) are indeed related to each other.

Moreover, the standard properties of the monopole current, as are observed
on the lattice, cannot be derived from (\ref{current}) alone but
imply also some non-local constraints on the Yang-Mills fields
\cite{brandt,strings}. Namely treating 
the continuum theory as a limiting case of the lattice gauge models 
one can argue \cite{strings} that:
\beq
j^a_{\mu}(x)~=~ \int d\tau \,\dot{\tilde{x}}_\mu\, 
n^a(\tau)\,\delta^4(x - \tilde{x}(\tau))
~=~ n^a(x)\, j_\mu(x)\,,
\eeq
where the colorless current $j_\mu$ is conserved, $\partial_\mu j_\mu ~=~ 0$, 
and $n^a(x)$ is determined in terms of the non-Abelian field strength tensor 
$G^a_{\mu\nu}$.

Moreover, it follows from Eq.~(\ref{current}) that the current $j_\mu^a$ satisfies the condition
\beq
D_{\mu}j_{\mu}^a~=~0\,,
\eeq
which means in turn that along the current trajectory the current
can be rotated to a certain direction in the color space:
\beq\label{cthree}
j^a_{\mu} ~ \to ~  \delta^{a3}\,j_{\mu}\,.
\eeq 
The whole construction was realized explicitly in case of a single
{\it external} magnetic current introduced via the the 't Hooft loop
and allowed for calculation of the heavy monopole potential,
with the gluon loop corrections included.

Also, Eq.~(\ref{cthree}) is the closest point to which theory can be brought
in attempt to accommodate the monopole dominance observed in the
lattice  simulations \cite{dominance}. Indeed, all the magnetic currents
can be aligned with the third direction in the color space.
However, the interaction of the currents is still sensitive,
generally speaking, to the ``charged'' gluons and the Abelian 
dominance remains empirical in nature.

To summarize, the attempt to define magnetic currents in the continuum
involves both singular fields, which are responsible for the violations
of the Bianchi identities, and non-local constraints which should
reflect the non-perturbative dynamics of the Yang-Mills theory
but at present cannot be checked independently.
Field theory is only applicable in cases when only one kind of charges
(magnetic or color) are dynamical while the dual charges are 
external (infinitely heavy). Indeed, only in this case the non-localities
might be avoided. On the other hand, it is clear that the consideration
of full charge-monopole dynamics goes beyond the local field theory context.

\section{Conclusions}

To summarize, we did not have much theoretical difficulties 
with the definition of mo\-no\-po\-les, or with the C-parity
of the corresponding current. 
Moreover the fine tuning observed at the presently available
lattices demonstrates once again the physical nature of the monopoles.
However, the description of the magnetic monopole dynamics in the continuum terms 
remains mostly unsolved problem. The fine tuning of the monopoles
exhibits their double-face nature.
Indeed, the monopoles are associated with  singular fields
and their formal definitions Eqs.~(\ref{bianchi},\ref{current})
look completely local. However, if this would be the case,
the infrared QCD scale, Eq.~(\ref{imaging}), could not appear.
And, indeed, in both cases the  definitions involve 
non-localities as well since
require either the gauge fixing over the whole lattice volume
or introduction of non-local constraints.
The intrinsic non-locality of the monopoles made also manifest, for example,
by the fact that their contribution to the vacuum energy
appears of order $\Lambda_{QCD}^4$, with no ultraviolet divergences
involved. 

The very fact that the dynamics of the monopoles is 
``complicated'' is of no surprise of course. Indeed, one 
searches for the monopoles in an attempt to identify 
degrees of freedom of the theory dual to the Yang-Mills theories.
Theory dual to the U(1) gauge theory is a field theory again.
But this case covers only the Abelian dominance (which cannot be exact~\cite{green}).
As for the SU(2) case there is no much hope 
that the dual theory is a  field theory again \cite{maldacena}.
On the positive side, one might say that the lattice monopoles have already
exhibited many unexpected and remarkable features which prove their
relevance to the confinement mechanism. Therefore, one could hope
that further phenomenological studies of the lattice monopoles
could provide us with a key to understand the structure of
the theory dual to the SU(2) gluodynamics.

\end{document}